\documentclass[
 reprint,
superscriptaddress,
amsmath,amssymb,
 aps,
pra,
]{revtex4-2}

\usepackage{graphicx}
\usepackage{dcolumn}
\usepackage{bm}

\begin{document}

\preprint{APS/123-QED}

\title{Observation of 2D Mott insulator and $\pi$-superfluid quantum phase transition in shaking optical lattice}

\author{Jingxin Sun}
\affiliation{School of Electronics, Peking University, Beijing 100871, China}

\author{Pengju Zhao}
\affiliation{School of Physics, Peking University, Beijing 100871, China}

\author{Zhongshu Hu}
\affiliation{School of Physics, Peking University, Beijing 100871, China}

\author{Shengjie Jin}
\affiliation{School of Physics, Peking University, Beijing 100871, China}

\author{Ren Liao}
\affiliation{School of Electronics, Peking University, Beijing 100871, China}

\author{Xiong-Jun Liu}
\email{xiongjunliu@pku.edu.cn}
\affiliation{School of Physics, Peking University, Beijing 100871, China}

\author{Xuzong Chen}
\email{xuzongchen@pku.edu.cn}
\affiliation{School of Electronics, Peking University, Beijing 100871, China}

\date{\today}

\begin{abstract}
The Mott insulator and superfluid phase transition is one of the most prominent phenomena in ultracold atoms. In this work, we report the observation of a novel 2D quantum phase transition between Mott insulator and $\pi$ superfluid in a shaking optical lattice. In the deep optical lattice regime, the lowest $s$-band can be tuned to Mott phase, while the higher $p_{x,y}$ bands are itinerant for having larger bandwidth. Through a shaking technique coupling the $s$ orbital to $p_{x,y}$ orbital states, we experimentally observe the transition between the states of the $s$ and $p_{x,y}$ bands, leading to a quantum phase transition from 2D $s$-orbital Mott phase to the $p_{x,y}$-orbital superfluid which condensed at $(\pi,\pi)$ momentum. 
By using the band mapping method, we also observe the changes of atomic population in different energy bands during the transition, and the experimental results are well consistent with theoretical expectations.

\end{abstract}

\maketitle


\section{Introduction}
The ultracold atoms loaded in optical lattices have provided a simple and controllable experimental platform for studying correlated quantum physics~\cite{jaksch1998cold,sherson2010single,tarruell2012creating}. A most prominent phenomenon discovered in optical lattice is the quantum phase transition from superfluid (SF) to Mott insulator (MI). Such SF-MI transition is based on the Bose-Hubbard model proposed by Fisher~\cite{fisher1989boson}. This novel quantum phase transition was first observed experimentally by M. Greiner et al. in 2002, signifying a milestone progress that the study of ultracold atoms turned from the weakly interacting regime to the strongly interacting regime~\cite{greiner2002quantum,greiner2002collapse}. The observation of SF-MI phase transition triggered a series of theoretical and experimental studies, such as many-body entanglement~\cite{briegel2001persistent,mandel2003controlled}, low-dimensional SF-MI transition~\cite{stoferle2004transition,spielman2008condensate}, short-range coherence in the Mott insulator~\cite{gerbier2005interference,gerbier2005phase} and fermionic Mott insulator~\cite{jordens2008mott,schneider2008metallic}.

While the lowest $s$ orbital physics have been mostly considered in optical lattice, the engineering of higher-orbital states has become feasible and attracted growing attention in the recent years~\cite{wirth2011evidence,lewenstein2011orbital,viebahn2021suppressing}. Due to the presence of the orbital degeneracy, the $p$-orbital optical lattice exhibit novel interacting effects~\cite{isacsson2005multiflavor,kuklov2006unconventional} including the spontaneous formation of orbital orders~\cite{liu2006atomic,PhysRevLett.100.200406,li2011bose,li2011effective,li2012time}. 
By applying a periodic frequency modulation to optical lattice, the coupling between $s$ and $p$ orbital states can be induced, allowing for loading and manipulation of atoms in the $p$-band. This makes it feasible to explore in experiment the rich many-body physics for atoms excited in the $p$ bands. In particular, the 1D gliding phase~\cite{li2016physics,morimoto2017floquet,ticco2016subharmonic}, 2D $p$-band Potts-nematic superfluidity~\cite{PhysRevLett.121.265301,jin2021evidence} and $p$-band superfluid with chiral topological phonon excitations~\cite{wang2021evidence} in a hexagonal optical lattice were observed through modulation of the lattices. 
Moreover, the $p$ orbital states can also be applied to simulate an effective strong ferromagnetic interaction is achieved, with the ferromagnetic-antiferromagnetic phase transitions and the formation of magnetic domains being also observed~\cite{clark2016universal,clark2018observation,sun2023quantitative,parker2013direct,sachdeva2017creating}.

When considering both the $s$-band and $p$-band simultaneously, the hybridization of these orbitals can be utilized to realize various forms of effective Hamiltonians~\cite{parker2013direct,ha2015roton}. Particularly, the characteristics of inter-orbital resonant coupling can be employed to achieve topological structures~\cite{liu2010quantum,li2013topological}. It was shown recently that the interactions between $s-p$ mixed orbital superfluids give rise to an unconventional type-II Nambu-Goldstone bosons with an emergent topological origin~\cite{pan2020emergence}. Additionally, the different orbitals can mimic different spin components~\cite{clark2018observation,gorg2019realization}. This provides an alternative way to simulate the synthetic spin-orbit (SO) coupling for ultracold atoms, which is the key ingredient in realizing the rich SO and topological physics~\cite{liu2009effect,lin2011spin,ando2013topological,zhou2013unconventional,zhang2014shaping,mei2014topological,wu2016realization,huang2016experimental,zhang2018spin,sun2018highly,cooper2019topological,lu2020ideal,yue2020enhanced,wang2021realization,lu2022floquet}. On the other hand, due to the significantly different bandwidths, the $s$-orbital and $p$-orbital can be modulated separately and are simultaneously tuned into the Mott insulator phase and superfluid phase, respectively. Based on this feature, a recent experimental work has reported a novel 1D quantum phase transition between $s$-band Mott insulator and $p$-orbital $\pi$-superfluid using Floquet modulation which couples the $s-p$ states~\cite{song2022realizing}. Unlike the previous Mott-superfluid phase transition in pure $s$ band, the 1D Mott-$\pi$SF transition is achieved by hybridizing two different obitals, i.e. the strongly interacting $s$-orbital and the weakly interacting itinerary $p$-band. 

In this experiment, we report the observation of the first 2D quantum phase transition between Mott insulator and $\pi$-superfluid by building a controllable Floquet modulation by shaking method to couple the $s$ orbital and $p_{x,y}$ orbitals in a square optical lattice. In the deep lattice regime, the $s$-band is tuned into Mott phase, while the $p_{x,y}$-bands are still itinerary. By shaking the optical lattice with frequency matching the energy difference between $s$ and $p$ orbitals, the hybridization between such two types of orbitals leads to the transition from Mott phase to superfluid one when the shaking strength is strong enough. The population dynamics of different bands are measured using band mapping techniques which provide a clear observation of the 2D Mott-$\pi$SF transition. The experimental results are well consistent with the theoretical calculations.

\section{Model description and experimental setup}

When the lattice is in the moving coordinate system, the one-dimensional equivalent Hamiltonian of the system can be expressed as:
\begin{equation}
	\label{1D hamiltonian}
	\begin{aligned}
		\hat{H}(t)=&\int dy\hat{\Psi }^{\dagger }(y)\left [ -\frac{\hbar ^2}{2m}\partial ^2y+V_{lattice}(y)+F(t)y  \right ]\hat{\Psi }(y) \\
		&+\frac{g_{1D}}{2m}
		\hat{\Psi }^{\dagger }(y)\hat{\Psi }^{\dagger }(y)\hat{\Psi }(y)\hat{\Psi }(y)
	\end{aligned}
\end{equation}
Where $\hat{\Psi}(y)$ is the boson field operator, m is the mass, $V_{lattice}$ is the lattice potential of the optical lattice before adding modulation, $g_{1D}$ is the equivalent interaction strength, $F(t)=m\ddot{s}(t)=m\omega^2Asin(2\pi ft)$ is the inertia generated by the optical lattice displacement s(t) caused by vibration the force, where $\omega=2\pi f$, which resonates with the energy gaps of the two lowest energy bands of Bloch, can couple these two energy bands.
In the experiment, what we changed was changing the lattice vibration frequency and vibration amplitude. So when other parameters are fixed and only the lattice vibration frequency $f$ and vibration amplitude $A$ are changed, we assume that the interaction is 0, and the equivalent Hamiltonian can be written as:
\begin{equation}
	\label{Hybridized Band Hamiltonian}
	\begin{aligned}
		\hat{H}=&\sum_k(-2J_s\cos k) \hat{s}_k^{\dagger}\hat{s}_k+ (2J_p\cos k+\Delta_p-hf) \hat{p}_k^{ \dagger}\hat{p}_k \\
		&+\frac{\Omega}{2} (\hat{p}_k^{\dagger}\hat{s}_k+H.c.)
	\end{aligned}
\end{equation}

When we ignore the next-nearest neighbor and longer-distance tunneling, the two terms of $J_s$ and $J_p$ can be approximated as the energy of S band and P band. The equation \ref{Hybridized Band Hamiltonian} can be understood as the translation between the P band and the S band through $\Omega$ term to form a hybridized energy band. $\Delta_p$ is the energy gap between the S-band and P-band.

\begin{figure}[htbp]
	\centering
	\includegraphics[width=\linewidth]{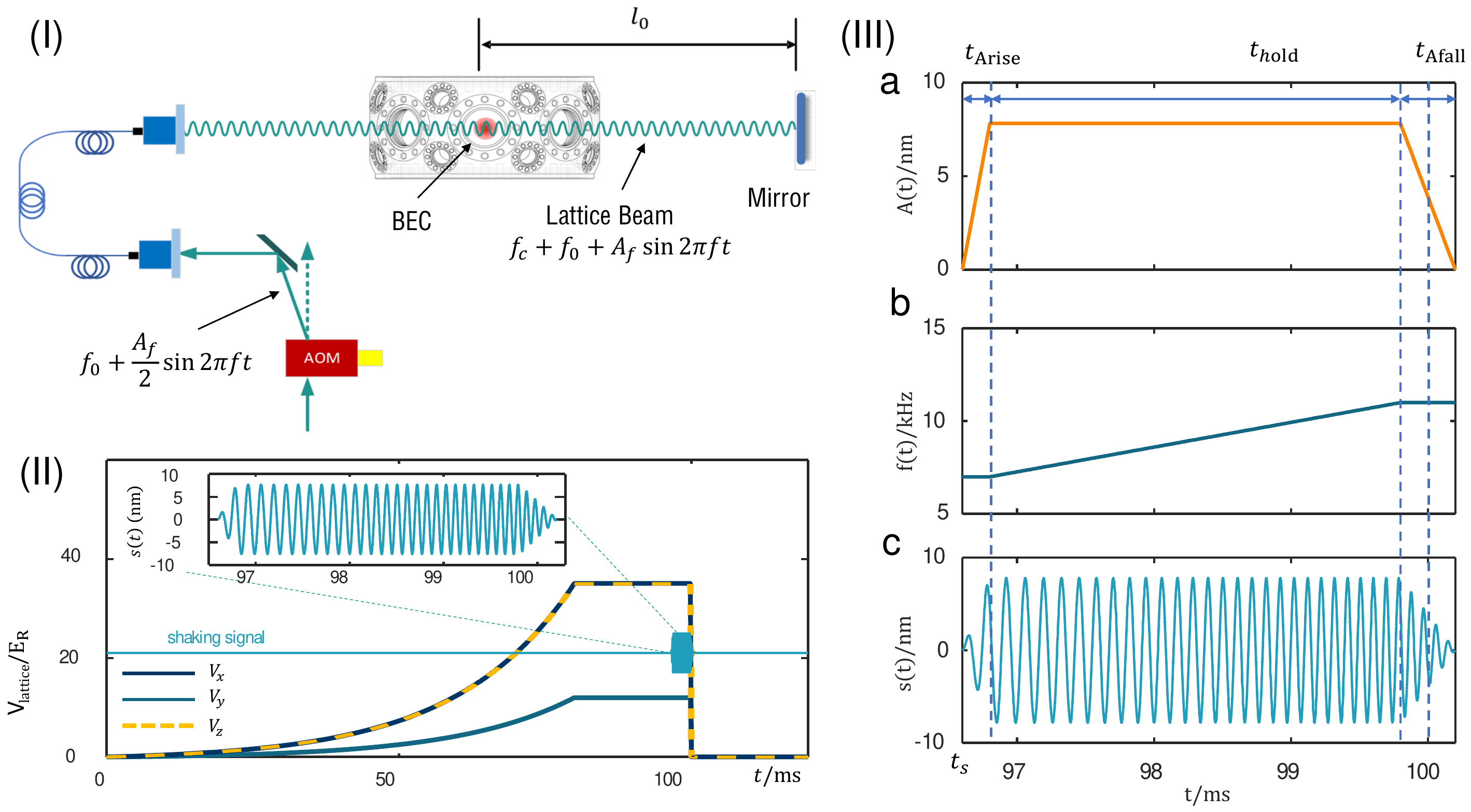}
	\caption{(I)Schematic diagram of the experimental setup used in the experiment. In the experiment, the AOM is used to shift the frequency of the optical lattice light by $f_0+A_fsin(2\pi ft)$, and then the input fiber is transmitted to the vicinity of the vacuum cavity, and is focused by the lens into the center of the vacuum chamber. After exiting the vacuum cavity, a zero-degree mirror is used to make the optical lattice The light returns in the original path, and the distance between the zero-degree mirror and the center of the chamber is $l_0$. (II) The loading timing of the optical lattice in the experiment. (III) Timing diagram of the optical lattice modulation signal. a. The modulation depth of the modulation signal is the amplitude A of the optical lattice shaking. b. The frequency f of the modulating signal. c. Displacement S of the optical lattice.}
	\label{fig1}
\end{figure}

Our experimental setup is shown in Fig.\ref{fig1}. The optical lattice is frequency-shifted by an acousto-optic modulator (AOM), and then the input fiber is transmitted to the vicinity of the vacuum cavity. We use double pass scheme in the experiment to ensure that the optical path is as unchanged as possible during the frequency modulation, thereby reducing the jitter of the optical lattice power during the modulation, so the frequency shift $f_0=2\times 110$ MHz. After being collimated and output from the optical fiber near the vacuum cavity, it is focused to the BEC position through a doublet lens with a focal length of 250 mm. The distance between the mirror and the center of the cavity is $l_0$.
The optical lattices in the X and Y directions have the same structure.

To verify whether the system can meet the experimental requirements in the optical lattice modulation experiment, we first modulate a one-dimensional optical lattice to determine the stability of the system. We use atoms in our experiments. We first obtain a BEC of $^{87}$Rb by evaporative cooling and the atoms number is $1\times 10^5$.Then we adiabatically load the optical lattice within 80ms with a time constant of 20ms. Among them, the final well depth in the x and z directions is $V_x=V_z=35$ Er, and the well depth in the y direction is $V_y=12$ Er. Then we keep the atoms in the optical lattice for 16.6ms to ensure that the atoms are in a balanced state. The moment when the optical lattice is loaded is the zero point of time. When $t_s=96.6$ ms, a modulation signal is added to the optical lattice in the y direction. After the modulation is finished, all potential fields are turned off, and the atoms are allowed to fall freely under the action of gravity. Absorption imaging was performed after a TOF of 30ms. The timing diagram of modulation signal loading is shown in Figure.\ref{fig1}(II) (III).

The modulation of the optical lattice is realized by adding a frequency modulation to the radio frequency signal of the optical lattice AOM, wherein the modulation frequency is $\frac{A_f}{2}sin(2\pi ft)$. The RF signal is provided by a high-precision voltage-controlled oscillator (POS-150+ minicircuits, VCO). A voltage control is given to the VCO by a function generator (33521B, Keysight) identical to the optical lattice control, corresponding to the modulation of the frequency of the VCO. $f_0$ is the frequency of AOM radio frequency source before adding modulation, $f_{c}$ is the frequency of optical lattice before adding modulation, $f$ is the modulation frequency, and $A_f$ is the amplitude of modulation frequency oscillation.
When the frequency of the optical lattice changes, the lattice will produce a shaking along the direction of the lattice, and the amplitude of the shaking can be calculated by the following method. The electric field of the optical lattice light is:
\begin{align}
	E(r,t)=E_0e^{-i\omega t}(e^{ikr}+e^{ik(2l_0-r)})
\end{align}
The light intensity of the light lattice is:
\begin{align}
	I(r,t)=4E_0^2\cos^2(kl_0-kr)
\end{align}
Where $k=k_c+\frac{2\pi A_f}{c} \sin(2\pi ft)=k_c+\delta k(t)$ is the light wave vector. $k_c$ is the light wave vector of optical lattice before adding modulation. Since $r$ is within the range of the light waist of the optical lattice, $I(r,t)\approx I_0\cos^2(k_cr-\delta k(t)l_0+\phi_0)$, where the phase part can be re-define the starting time point to eliminate, then the longitudinal vibration range of the peak and valley of the frequency vibration is:
\begin{align}
	s(t)=\frac{\delta k(t)l_0}{k_c}=\frac{\lambda l_0 A_f}{c}\sin(2\pi ft)=A\sin(2\pi ft)
\end{align}
In our experiment, $\lambda=1064\mathrm{nm}, l_0=40$ cm, it can be estimated that when $A_f=1$ MHz, the amplitude of the corresponding lattice vibration is $A=1.4$ nm. In the optical lattice loading stage, we maintain the modulation signal as $V_0=8$ V, corresponding to the center frequency of the AOM in the y direction as 110MHz. Then when the optical lattice is maintained in the optical trap for a period of time, when $t_s=96.6$ms, a modulation signal $A_fsin(2\pi ft)$is added to the control signal. Loading the oscillation amplitude of the modulating signal from 0 to $A_f$ within $t_{Arise}=0.2$ ms. At this time, the frequency of the modulating signal is $f_0$ and then within $t_{hold}=3$ms, linearly load the oscillation frequency from $f_0$ to $f_e$, here During the process, the oscillation amplitude remains unchanged at $A_f$. Finally, reduce the amplitude from $A_f$ to 0 within $t_{fall}=0.4$ ms. The atoms are then imaged after TOF with all potential fields turned off.

\section{Observation of 1D MOTT-$\pi$SF phase transition}%

\begin{figure}[hbp]
	\centering
	\includegraphics[width=\linewidth]{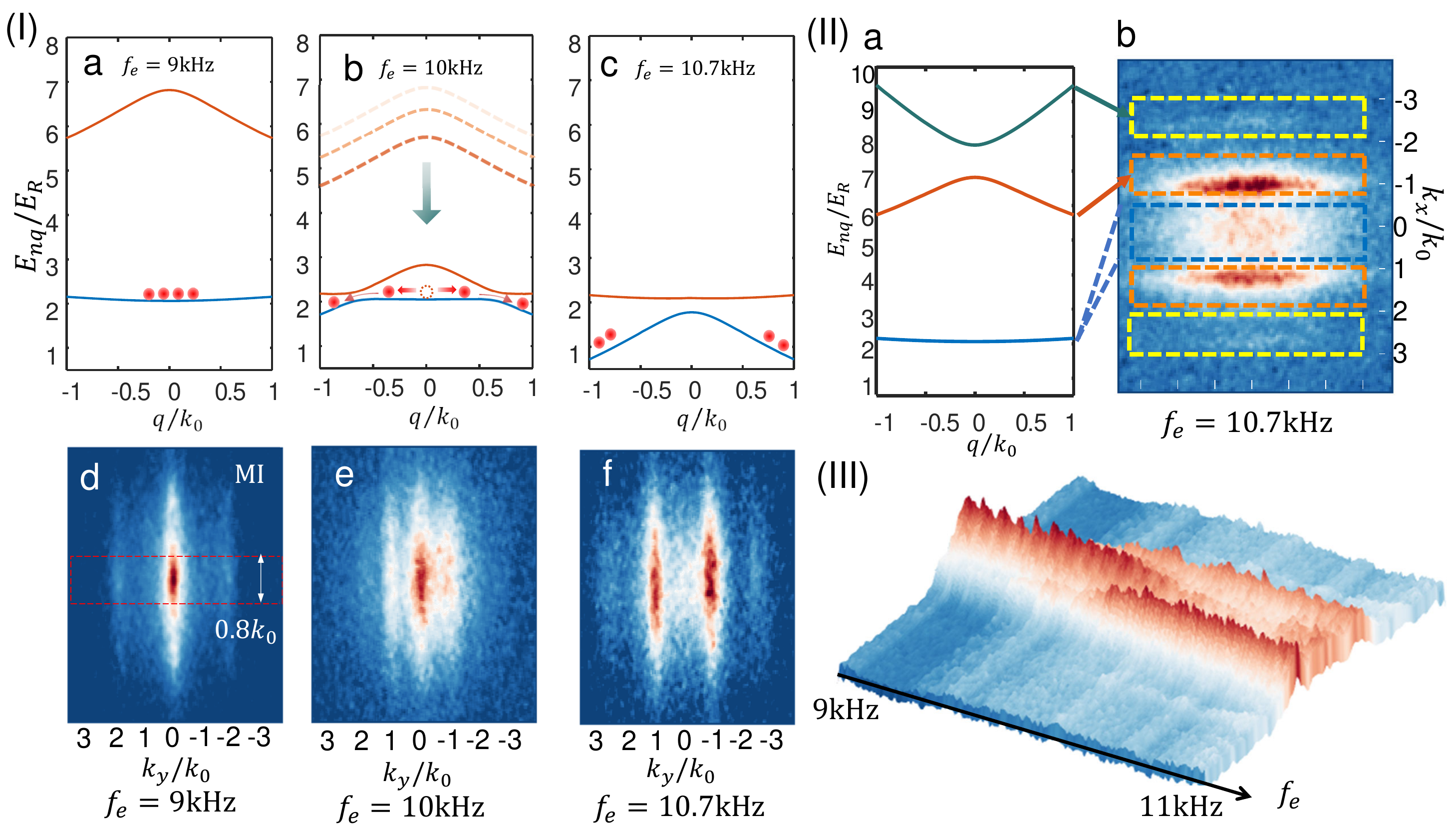}
	\caption{1D Mott-$\pi$SF phase transition process. (I) a. Bloch band distribution of the 1D Mott state without phase transition (blue line for S band, red line for P band). b. When the optical lattice modulation frequency is $f_e=10$ kHz, the S band and P band begin to couple, and the atoms gradually condense at the lowest point of the coupled band, located at $k/k_0=\pm1$. c. As the modulation frequency increases to $f_e=10.7$ kHz, band inversion occurs between the S band and P band, and the atoms gather at $k/k_0=\pm1$, transitioning from the Mott state to the $\pi$SF. Correspondingly, the atoms are imaged using the TOF method at different modulation frequencies. d. $f_e=9$ kHz, e. $f_e=10$ kHz, f. $f_e=10.7$ kHz. (II) Correspondence between the 1D optical lattice Bloch band and the observed atomic distribution using the bandmapping method. (III) Variation of the observed atomic distribution using the bandmapping method with the modulation frequency changing from $f_e=9$ kHz to $f_e=11$ kHz.}
	\label{fig2}
\end{figure}

Figure.\ref{fig2}.(I) shows the Bloch energy bands of a one-dimensional modulated optical lattice obtained from theoretical simulations. When $f_e=9$ kHz, the ground state is a Mott insulator where all atoms are in the S band. As $f_e$ gradually increases, the P band of the Bloch band structure shifts down and hybridizes with the S band, resulting in a complex ground state. When $f_e$ continues to increase, the two bands flip and become independent, and all atoms are located in the P band, forming a $\pi$ superfluid phase. The position of the lowest point of the hybridized band shifts from $q=0$ to $q=k_0$, where $k_0=2\pi /\lambda$, and $\lambda$ is the wavelength of the optical lattice, which is $1064$ nm in the experiment.

Atoms are distributed in the lowest point of the energy bands in momentum space. As the modulation frequency increases, the number of atoms located at $k_y/k_0=0$ gradually slides to both sides. When the modulation frequency reaches $f_e=10.7$ kHz, the atoms aggregate at $k_y/k_0=1$. The atom number distribution after time-of-flight (TOF) is shown in Fig.\ref{fig2}.(I)(d)-(f), which clearly shows the gradual change of atom distribution to $k_y/k_0=1$. This change is the quantum phase transition from the Mott insulator to the $\pi$ superfluid phase.

In the experimental TOF diagram, we only considered the coupling between the S and P bands. In reality, there is also coupling between the P and D bands. When most of the atoms are transferred to the P band, some of the atoms will also be transferred to the D band due to the coupling between the energy bands. However, because it is difficult to distinguish between the atoms in the D and S bands in the momentum distribution plot, we cannot see the change in the number of atoms in the D band very intuitively in the TOF image.

In order to observe the distribution of atoms in the S, P, and D bands more intuitively, we use the bandmapping method to image the atoms. Bandmapping is a method that can help us determine the energy band in which a particle is located in an experiment. By adiabatically turning off the optical lattice, the Bloch function will evolve into a free electron wave function. The wave vector of the free electron plane wave is the same as the quasi-momentum wave vector of the Bloch function. Because the energy is almost unchanged during this process, the Bloch function of the nth energy band will be mapped to a plane wave with a wave vector in the nth Brillouin zone. In this case, we can infer the energy band in which the particle originally resided by observing the momentum of the plane wave. In the experimental operation, after the modulation of the optical lattice is completed, we adiabatically reduce the depth $V_y$ of the y-direction optical lattice from $V_y=12$ Er to 0 Er within 2 ms. The x,z direction optical lattices are still quickly turned off to 0 Er. After 30 ms of TOF, the atoms are imaged by absorption, and the imaging of the atoms using the band mapping method can be obtained, as shown in Figure.\ref{fig2}.(II)b.

In Figure.\ref{fig2}.(II), the corresponding optical lattice Bloch energy bands of the different momentum ranges of the atomic images can be seen. The blue region corresponds to the distribution of atoms in the S band, the orange region corresponds to the distribution of atoms in the P band, and the yellow region corresponds to the distribution of atoms in the D band. Taking the atoms at the center of the x-direction in Figure.\ref{fig2}.(II).b with a momentum of $0.8k_x$, the variation of the momentum distribution of the atoms with $f_e$ can be observed, as shown in Figure.\ref{fig2}.(III). As the modulation frequency $f_e$ changes, the atoms gradually transfer from the S band to the P band, and when most of the atoms transfer to the P band, some of the atoms begin to transfer to the D band due to the coupling between the energy bands. It can be seen that the atomic distribution gradually transfers from S band to P and D bands.

\section{Observation of 2D MOTT-$\pi$SF phase transition}
When the reliability of the system is verified in a one-dimensional optical lattice, we extend the Mott-$\pi$SF phase transition of one-dimensional optical lattice to two-dimensional. At this time, the Hamiltonian of the system can be expressed as:

\begin{align}
	\hat{H}=\hat{H}_x+\hat{H}_y
\end{align}

Similar to the modulation of the one-dimensional optical lattice, after obtaining the BEC, we adiabatically load the optical lattice within 80 ms with a time constant of 20 ms. The final trap depth in the z direction remains $V_z=35$ Er, and the trap depth in the x and y directions is $V_x=V_y=12$ Er. Then we maintain the atoms in the optical lattice for 16.4 ms to ensure that they are in the equilibrium state. Similarly, taking the time of optical lattice loading as the time zero point, when $t_s=96.4$ ms, we apply modulation signals to the optical lattice in the x and y directions.

\begin{figure}[b]
	\centering
	\includegraphics[width=\linewidth]{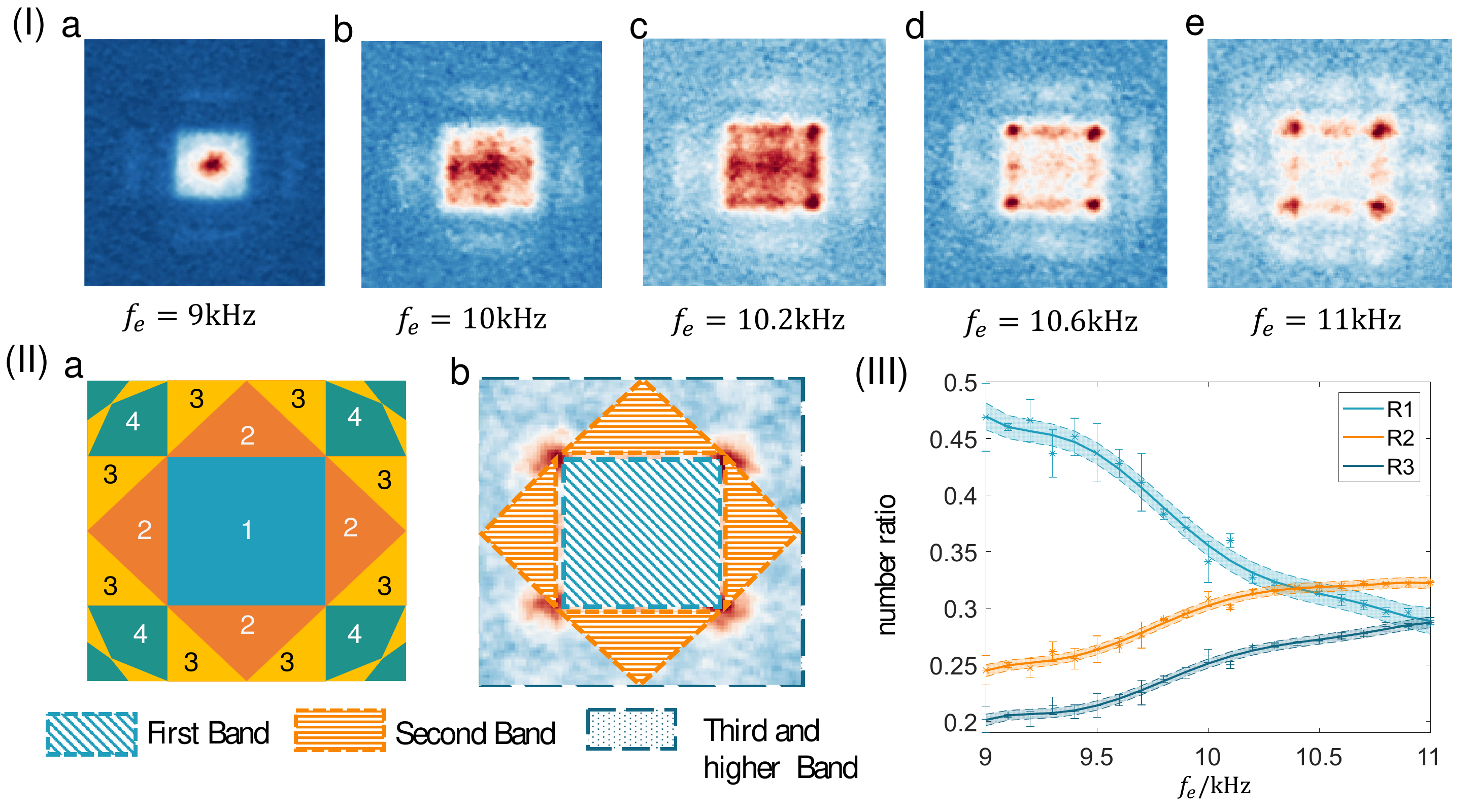}
	\caption{Observation of atomic distribution in 2D Mott-$\pi$SF phase transition using bandmapping method. (I) Atomic number distribution of the 2D optical lattice Mott-$\pi$SF using bandmapping at different modulation frequencies in x and y directions. (II) a. Reciprocal lattice and Brillouin zones for a 2D lattice. b. Correspondence between the atomic distribution measured by the bandmapping method and the Brillouin zone of the 2D lattice. (III) atomic population of the first band, second band, and higher bands of the optical lattice as a function of modulation frequency $f_e$. The green $R1$ corresponds to the atomic population of the first band, orange $R2$ corresponds to the second band, and dark blue $R3$ corresponds to higher energy band atomic population.}
	\label{fig3}
\end{figure}

In 1D experiment, we observed that the bandmapping method can better observe the population distribution of atoms in different energy bands, making it easier to compare with theoretical simulations. Therefore, in the two-dimensional lattice experiment, we use the bandmapping imaging method. After the modulation of the lattice ends, we adiabatically reduce the depths of the optical lattice in the x and y directions, $V_x$ and $V_y$, from $V_x=V_y=12$ Er to 0 Er within 2ms. The optical lattice in the z direction is still quickly closed to 0 Er. After 30 ms of TOF, we perform absorption imaging on the atoms. In Fig.\ref{fig3}(I), we can observe that as the modulation frequency $f_e$ varies from 9 kHz to 11 kHz, the distribution of atom numbers gradually moves from the central region of $(k_x,k_y)$ in the four points of $(\pm k_0, \pm k_0)$ to other regions.

Since we used the bandmapping method, similar to the one-dimensional experiment, the distribution of the two-dimensional atom numbers in the imaging results can correspond to the Brillouin zone of the two-dimensional optical lattice. As shown in Fig.\ref{fig3}.(II), the green diagonal line corresponds to the first Brillouin zone, the orange horizontal line corresponds to the second Brillouin zone, and the blue dots correspond to the third and fourth Brillouin zones. The ratio of the number of atoms in different Brillouin zones to the total number of atoms is taken as the population of atoms in that energy band. We can see that the population of atoms in different energy bands changes with the modulation frequency of the optical lattice. We can observe the transition from the Mott state to the $\pi$SF state, and the phase transition point is around $f_e=10.3$ kHz.

\begin{figure}[t]
	\centering
	\includegraphics[width=\linewidth]{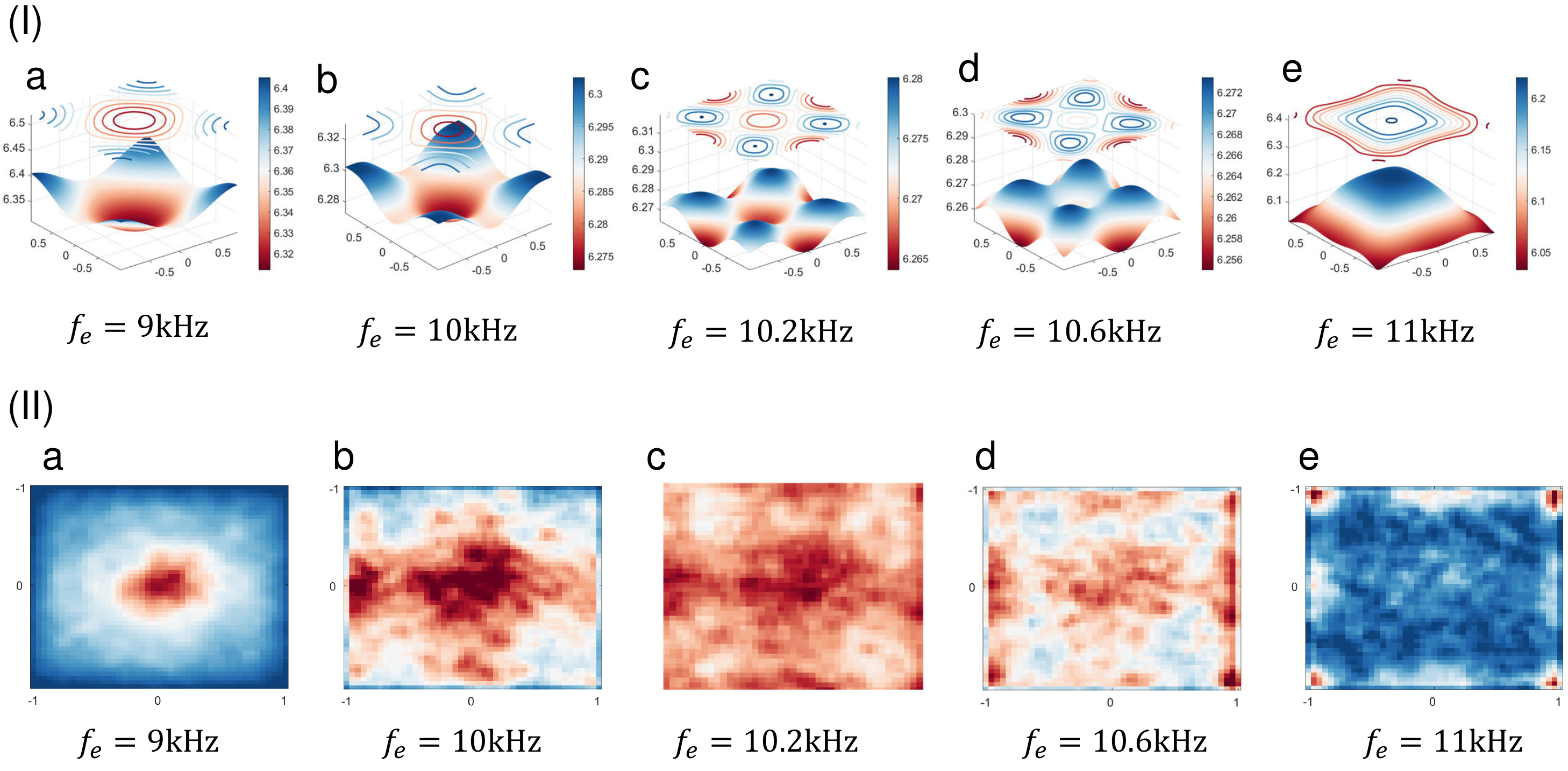}
	\caption{Two-dimensional Mott-$\pi$SF phase transition. (I) Theoretical simulation of the lowest energy band of a 2D optical lattice after coupling at different modulation frequencies. (II) Experimental images of the first Brillouin zone of the 2D optical lattice at different modulation frequencies.}
	\label{fig4}
\end{figure}

Similar to the one-dimensional optical lattice, we can theoretically simulate the Bloch bands of a two-dimensional optical lattice. As shown in Figure.\ref{fig4}.(I), it is a theoretical simulation of the momentum distribution of the Bloch band S of a two-dimensional optical lattice. The first Brillouin zone in the Brillouin zone corresponds to the atomic momentum distribution of the Bloch band S. When we use the bandmapping method to perform absorption imaging observation of the atoms, similar to the one-dimensional optical lattice, in the imaging image, we can see a square momentum distribution with a width of $k/k_0=2$, which coincides with the first Brillouin zone. In Figure.\ref{fig3}. (I).a, we can see that the atoms are evenly distributed in the first Brillouin zone, and no atom population is detected in higher Brillouin zones. This proves that the atoms occupy the lowest energy band and no higher band is filled.

We compare the first Brillouin zone with the ground state of the theoretically simulated two-dimensional shaken optical lattice. As shown in Figure.\ref{fig4}.(I)(II), as the modulation frequency increases from $f_e=9$ kHz to $f_e=11$ kHz, the atomic distribution gradually moves from $(k_x,k_y)$ at position (0,0) to the four points of $(-k_0, k_0)$, $(-k_0,-k_0)$, $(k_0,-k_0)$, and $(k_0,k_0)$. The data results are consistent with the theoretical simulation.

\section{Conclusion and outlook}

In this work, we have experimentally studied a 2D quantum phase transition in square optical lattice by modulating the lattice shaking which couples the $s$ orbital and $p_{x,y}$ orbital states. With the band mapping technique which measures the population of atoms in different bands, we observed a novel phase transition from the $s$ orbital Mott insulator to $p$ orbital superfluid through increasing the shaking induced $s-p$ hybridization, with $p$-orbital superfluid being condensed at the $(\pi,\pi)$ momentum. The observed Mott-$\pi$SF phase transition agrees well with theoretical simulation. 

The Mott-$\pi$SF phase transition in shaking optical lattices is a novel new research topic, in which the quantum many-body physics are enriched due to the orbital degree of freedom. Future interesting issues including to explore the fundamental physical mechanisms of this phase transition, including the critical phenomena, quantum coherence, and quantum dynamics. In addition, one may study this phase transition in other systems, such as in topological quantum matter, quantum computing and quantum simulation systems. 
Moreover, in general the orbital ordering can be achieved in the $p$-band. The transition between $s$-band Mott insulator and orbital ordered $\pi$-superfluid is a novel issue deserving the next study. Finally, the coupling between $s$ band and even higher bands may also bring rich quantum many-body physics. 

This work is supported by the National Natural Science Foundation of China (Grants No. 11920101004, 11934002), and the National Key Research and Development Program of China (Grant No. 2021YFA1400900, 2021YFA0718300).

\bibliography{2DMOTT-PiSF}

\end{document}